# Kondo scattering in underdoped Nd$_{1-x}$Sr$_x$NiO$_2$ infinite-layer superconducting thin films

T. N. Shao[1], Z. T. Zhang[1], Y. J. Qiao[1], Q. Zhao[1], H. W. Liu[1*], X. X. Chen[1], W. M. Jiang[1], C. L. Yao[1], X. Y. Chen[1], M. H. Chen[1], R. F. Dou[1], C. M. Xiong[1], G. M. Zhang[2,3*], Y.-F. Yang[4,5,6*], J. C. Nie[1*]

[1]*Department of Physics, Beijing Normal University, Beijing 100875, China*

[2]*State Key Laboratory of Low-Dimensional Quantum Physics and Department of Physics, Tsinghua University, Beijing 100084, China*

[3]*Frontier Science Center for Quantum Information, Beijing 100084, China*

[4]*Beijing National Laboratory for Condensed Matter Physics and Institute of Physics, Chinese Academy of Sciences, Beijing 100190, China*

[5]*School of Physical Sciences, University of Chinese Academy of Sciences, Beijing 100190, China*

[6]*Songshan Lake Materials Laboratory, Dongguan, Guangdong 523808, China*

*Corresponding author. E-mail: jcnie@bnu.edu.cn (J.C.N.); yifeng@iphy.ac.cn (Y.-F.Y.); gmzhang@tsinghua.edu.cn(G.M.Z.); haiwen.liu@bnu.edu.cn (H.W.L.)

**The recent discovery of superconductivity in infinite-layer nickelates generates tremendous research endeavors, but the ground state of their parent compounds is still under debate. Here, we report experimental evidences for the dominant role of Kondo scattering in the underdoped Nd$_{1-x}$Sr$_x$NiO$_2$ thin films. A resistivity minimum associated with logarithmic temperature dependence in both longitudinal and Hall resistivities are observed in the underdoped Nd$_{1-x}$Sr$_x$NiO$_2$ samples before the superconducting transition. A linear scaling behavior $\sigma_{xy}^{AHE} \sim \sigma_{xx}$ between anomalous Hall conductivity $\sigma_{xy}^{AHE}$ and conductivity $\sigma_{xx}$ is revealed, verifying the dominant Kondo scattering at low temperature. The effect of weak (anti-)localization is found to be secondary. Our experiments can help clarifying the basic physics in the underdoped Nd$_{1-x}$Sr$_x$NiO$_2$ infinite-layer thin films.**

The mechanism of high-T$_c$ superconductivity remains a long-standing mystery, even though tremendous efforts and significant advances have been made in the study of cuprates and Fe-based superconductors. Aiming to mimic the cuprate-like electronic configuration, superconductivity has been proposed and recently found in the infinite-layer nickelate compounds [1]. This discovery has drawn enormous attention [2-6] since it may provide deeper insight about the pairing mechanism of unconventional superconductivity. But instead of an antiferromagnetic (AFM) order in the cuprates, the parent compound of superconducting nickelates shows no sign of static long-range magnetic order down to 1.7 K by neutron diffraction [7] and down to 2 K by muon spin measurements [8] in bulks and also a lack of long-range AFM in thin films [9], despite their similarities in the crystalline and electronic structures with the formal $3d^9$ configuration. Moreover, unlike the cuprates where doped holes reside on the oxygen atoms, here they are mainly introduced into Ni 3d states and reside in the $3d_{x^2-y^2}$ orbitals [10], supported by the softening of resonant inelastic x-ray scattering (RIXS) [11]. Besides, superconductivity remains absent in infinite-layer nickelate bulk, although the application of pressure up to 50.2 GPa can significantly suppress the insulating behavior [12,13]. This raises a question regarding whether the cuprates and hole doped $NdNiO_2$ share the same mechanism of superconductivity [14]. In the experimental aspect, controlling the doping concentration and studying its relationship with $T_c$ has become a necessary step to understand the electron interaction in nickelate systems. A non-monotonic relationship between $T_c$ and Sr doping concentration is established [3,15], reminiscent of the competition between superconductivity and adjacent phase (or its phase fluctuation) in unconventional superconductors [16-18]. It is therefore important to understand the underdoped nickelates, where, different from the cuprates, the resistivity shows metallicity at high temperatures and weak insulating behavior at low temperatures.

In order to explain the resistivity upturn at low temperatures, Zhang et al. [19] proposed a self-doped Mott-Kondo scenario for the parent nickelate system, bridging the Kondo lattice model for heavy fermions and the t-J model for cuprates. It was proposed that self-doping, namely charge transfer from localized Ni 3d orbitals to other conduction bands, plays an essential role. The remaining Ni 3d local moments may couple to the conduction electrons, causing the well-known Kondo screening physics [20] and giving rise to Kondo scattering that explains the low temperature resistivity upturn reported in $NdNiO_2$ [1], $LaNiO_2$ [9], as well as underdoped infinite-layer $Nd_{1-x}Sr_xNiO_2$ (x = 0.1, 0.125) [15]. The absence of long-range magnetic ordering [9] might also be attributed to the self-doping and the Kondo screening effect. However, in addition to other theoretical scenarios including the pseudogap [21], magnetic scattering [22] and the d-wave order [23], the logarithmic temperature dependence of resistivity at low temperatures has also been observed in underdoped cuprates and was mainly attributed to the weak localization (WL) [24-26]. Although it has been argued theoretically [5,27] that the transport properties of the underdoped nickelates are very different from that of the AFM cuprate Mott insulator, whether the *lnT* behavior in the underdoped nickelates is caused by the Kondo effect or by the WL effect has not been determined, which leads to unsettled debate concerning the basic physics of the nickelate superconductors. The transport measurements including both longitudinal and transverse resistivity are therefore indispensable to uncover the low-energy excitations of the parent compounds of nickelate superconductor and provide decisive evidence for different theoretical proposals to understand the unconventional superconductivity in nickelate system.

Here, we study the normal-state transport properties of the infinite-layer $Nd_{1-x}Sr_xNiO_2$ thin films with a low Sr doping concentration. Our preliminary analysis of the film reveals a *lnT* behavior in both resistivity and $R_H$ in the same temperature region. A linear dependence of the anomalous Hall-effect (AHE) conductance $\sigma_{xy}^{AHE} \sim \sigma_{xx}$ indicates that the skew scattering is dominant in the exact same temperature range. The dephasing rate shows a linear temperature dependence at temperatures below $T_K$, in good agreement with the Kondo scenario. The WL contribution is shown to be secondary. Our experimental results strongly support the self-doped Mott-Kondo scenario for the underdoped $Nd_{1-x}Sr_xNiO_2$ infinite-layer superconducting thin films.

***lnT* resistance.** — Figure 1a shows the temperature-dependent resistivity $\rho(T)$ for the underdoped $Nd_{0.88}Sr_{0.12}NiO_2$ thin film. A superconducting transition is observed, with an onset at 4.06 K and a midpoint at 0.79 K. The broad transition indicates the inhomogeneity of the infinite-layer thin film. These observations reveal that the infinite-layer nickelate phase (the structural characterizations are shown in Supplementary Figs. 1-2), not the reduced secondary phase, is superconducting. As shown in Fig. 1a, the resistivity of the $Nd_{0.88}Sr_{0.12}NiO_2$ thin film first decreases as the temperature decreases, followed by a resistivity minimum at a characteristic temperature $T^* \sim$ 40 K. The minimum resistivity (~0.80 mΩ·cm at 40 K) falls below the value of 0.87 mΩ·cm that corresponds to the quantum sheet resistance ($h/e^2 \sim$ 26 kΩ) per $NiO_2$ two-dimensional plane, which is consistent with the previous report [15]. Below 30 K down to 7 K, the resistivity shows logarithmic temperature dependence regardless of the magnetic field (see Fig. 1b). In fact, this characteristic is commonly observed in all our underdoped $Nd_{1-x}Sr_xNiO_2$ infinite-layer thin films, as shown in Fig. 1c. We find that the $\rho$-$T$ curves can be well described by Hamann model from 8 K to 120 K [28],

$$\rho(T) = \rho_0 + aT^2 + bT^5 + \rho_K(T/T_K) \tag{1}$$

where $\rho_0$ is the residual resistivity caused by sample disorder and the $T^2$ and $T^5$ terms are the contributions of electron-electron and phonon-electron interactions, respectively. $\rho_K(T/T_K)$ is the resistivity induced by magnetic scattering in the absence of magnetic field [28-30] and takes the form,

$$\rho_K(T/T_K) = \frac{2\pi c\hbar}{ne^2 k_F}\{1 - ln(T/T_K) \cdot [ln^2(T/T_K) + s(s+1)\pi^2]^{-1/2}\} \qquad (2)$$

where $k_F = 2\pi\left(\frac{3n}{8\pi}\right)^{1/3}$ is the Fermi wave-vector in a free electron approximation. The best fitting results to the $\rho$-T curves of underdoped Nd$_{1-x}$Sr$_x$NiO$_2$ samples using Eqs. (1) and (2) are shown in Supplementary Fig. 5. Moreover, we have carried out low-temperature transport measurements in magnetic field for three underdoped Nd$_{1-x}$Sr$_x$NiO$_2$ samples (x = 0, 0.05, 0.09). All three underdoped samples are well consistent with the Kondo scattering scenario [20] down to 0.04 K. The numerical renormalization group (NRG) fitting, non-crossing approximation (NCA) fitting (see Fig. 1d and Supplementary Fig. 4) and Hamann fitting (Supplementary Fig. 5) curves provide quantitative justification for this point. The magnetic field truly influences the *R-T* curves of underdoped samples with pronounced negative magnetoresistance (as shown in Supplementary Fig. 5), reminiscent of the well-known negative magnetoresistance in Kondo system (La, Ce)Al$_2$ [31]. Moreover, all the underdoped samples (Fig. 1d) demonstrate clear feature of Kondo scattering in temperature regime above $T_K$. The good agreement with the NRG, NCA, and Hamann predictions proves exclusively the Kondo mechanism and leaves little room for other interpretation in underdoped region, which also confirms that the local moment is roughly spin-1/2. This excludes its possible origin from the Nd 4f spin (S=3/2) and supports its origin from the Ni $3d_{x^2-y^2}$ moment.

**Pinpoint the Kondo scattering mechanism.**—Remarkably, for temperatures below 40 K down to 6 K, the $R_H$ follows closely that of the resistivity $\rho$, and resembles the *lnT* dependence of the resistivity, as shown in Fig. 2a. The $R_H \propto \rho$ behavior is well consistent with the theoretical prediction of Kondo skew scattering associated with local moments [32,33]. Thus, both the resistivity and Hall coefficient support the presence of the magnetic Kondo scattering in the underdoped nickelate superconductor. The relation between $\sigma_{xy}^{AHE}$ and $\sigma_{xx}$ can further confirm the skew scattering mechanism characterized by $\sigma_{xy}^{AHE} \propto \sigma_{xx}$ in a temperature range of 7–40 K for the Nd$_{0.88}$Sr$_{0.12}$NiO$_2$ thin film (see Fig. 2b). The conductivity $\sigma_{xx}$ of the Nd$_{0.88}$Sr$_{0.12}$NiO$_2$ is about 10$^3$ Ω$^{-1}$cm$^{-1}$ and is in a bad metal regime [33], in which $\sigma_{xy}^{AHE}$ should generally decrease with decreasing $\sigma_{xx}$ at a rate faster than linear. Nevertheless, by subtracting the ordinary Hall effect (OHE) contribution ($R_0$, which is determined by the measured Hall coefficients and the Curie-Weiss fit and does not change with varying temperature, see detailed discussion in Eqs. (4) and (5) and results in Fig. 3a) to obtain $\sigma_{xy}^{AHE}$ from $\sigma_{xy}$, a linear dependence of $\sigma_{xy}^{AHE} \propto \sigma_{xx}$ is observed in the temperature range of 7–40 K. This temperature range is the same as the range where resistivity shows logarithmic temperature dependence (see Fig. 2a). These observations strongly support that the incoherent skew scattering [32,33] is a dominant mechanism for understanding our $R_H(T)$ results. At high temperatures (> 40 K), the Kondo scattering is suppressed and a deviation from the linear dependence between $\sigma_{xy}^{AHE}$ and $\sigma_{xx}$ is also observed.

One may argue that the logarithmic temperature dependence of resistivity at low temperatures, a hallmark of the Kondo effect, could also originate from the weak localization/weak anti-localization (WL/WAL) in two-dimensional system. However, as shown in Fig. 1c, the *lnT* correction of resistivity under large magnetic field can exclude the WL/WAL correction and justify the Kondo scattering scenario. The conductance correction due to the Kondo effect can be obtained with $\Delta\sigma_{Kondo} = \sigma(T) - \sigma(T_{min} = 40K)$ (zero field) and the data (extracted from Fig. 1b) are shown in Fig. 3a. The red line is the $e^2/\pi h \cdot lnT$ fit for $\Delta\sigma$, the obtained slope is $\beta \approx 6.96$ for the temperature range of 8-24 K. Although both the WL/WAL and the electron-electron interaction (EEI) in two-dimensional system also contribute to a conductance correction proportional to $e^2/\pi h \cdot lnT$, the total value for the coefficient of WL/WAL and EEI corrections is commonly less than 2 [34]. Thus, the value of $\beta$ is much larger than the typical value contributed from WL/WAL and EEI effect (see the Supplementary Note 1) [30]. In order to single out the WL/WAL correction, the measurement in a modest magnetic field (*e.g.*, H = 4 T) is necessary (Supplementary Fig. 8) [30]. The conduction correction due to the WL/WAL effect can be obtained with $\Delta\sigma_{WL/WAL} = \sigma(T)|_{H=0} - \sigma(T)|_{H=4T}$, and the data are shown in the inset of Fig. 2c. The blue line is the $lnT$ fit, the obtained slope is $\Delta\kappa = \alpha p \approx 0.213$ ($p$ is the dephasing exponent, $\tau_\varphi^{-1} \sim T^p$) for the temperature range of 10–20 K, much less than the total value $\beta \approx 6.96$, indicating that the WL/WAL is a secondary effect and Kondo scattering is dominant in the temperature range of 7–30 K.

Moreover, we can also utilize the low field magneto-conductivity to analyze the temperature dependence of electron dephasing, based on the modified Hikami-Larkin-Nagaoka (HLN) formula [35]:

$$\Delta\sigma(H) = \frac{\alpha e^2}{\pi h}\left[\Psi\left(\frac{1}{2} + \frac{H_\varphi}{H}\right) - ln\frac{H_\varphi}{H}\right], \quad (3)$$

where $\Psi$ is the digamma function, $H_\varphi = \hbar/4eD\tau_\varphi$, $D$ is the electronic diffusion constant, and $\alpha$ is an effective constant depending on the relative strengths of magnetic scattering and spin-orbit coupling (See

Supplementary Figs. 8-9 for related results and see Supplementary Figs. 10-15 for detailed analysis) [30]. As shown in the inset of Fig. 2d, for the nickelate thin film studied in this work, a small positive value $\alpha \approx 0.20$ can be maintained for a wide temperature range of 9–20 K. As the dephasing rate $\tau_\varphi^{-1}$ is simply proportional to $H_\varphi$, there exists a linear power-law dependence: $H_\varphi \sim T^p$ with $p \approx 1$. Indeed, as shown in Fig. 2d, a linear temperature dependence of the dephasing field $H_\varphi$ is observed over a wide temperature range of 8–16 K below $T_K$, which is consistent with the universal dephasing rate due to diluted Kondo impurities [36]. The dephasing field $H_\varphi$ exhibits a deviation from the linearity around ~ 16.0 K. This is another definition of $T_K$ [37,38], and the obtained value (16.0 K) is well consistent with that (~15.6 K, see Supplementary Fig.5d) determined by the resistivity measurements. Thus, the Kondo scenario provides a consistent explanation of all our measured data.

**Phase diagram.**—Figure 3a displays the Hall-effect measurements obtained in large temperature range under an applied field of 9 T, showing negative Hall coefficients ($R_H$) with a maximum at $T^* \sim 40$ K. Interestingly, we find the Hall coefficients now follow a simple Curie-Weiss law. To see this, we separate the normal coefficient $R_0$ from the AHE coefficient and make the ansatz based on the treatment of heavy fermion superconductors [39]

$$\rho_{xy} = R_0 H + 4\pi M R_S. \tag{4}$$

Taking $M = \chi H$, $\chi = C/(T-\Theta)$, we have

$$R_H = \frac{\rho_{xy}}{H} = R_0 + 4\pi \frac{C}{T-\Theta} R_S = R_0 + \frac{R_S'}{T-\Theta}, \tag{5}$$

with three fitting parameters. $R_0$ is the OHE coefficient due to the deflection of the conduction electrons by the Lorentz force. We obtain a good fit which satisfies the Curie-Wiess law for 60 K $\leq T \leq$ 300 K. The best fit (solid line in Fig. 3a) was obtained at $R_0 = -2.61 \times 10^{-3}$ cm$^3$C$^{-1}$, $\Theta = -66.45$ K and $R_S' = 0.20$ cm$^3$KC$^{-1}$. $R_0$ was found to be negative, which means that the OHE is dominated by electrons. This is consistent with the band structure calculations revealing that the parent NdNiO$_2$ contains small electron pockets at the Fermi energy [40]. The fitting value $R_0$ corresponds to 0.12 electron per formula unit. As shown in Fig. 3b, the Hall resistivity $\rho_{xy}$ versus magnetic field up to 35 T at temperatures from 10 K to 70 K shows no obvious deviation from linear dependence on magnetic field (also see Supplementary Fig. 7 for the extended temperature region data). Such Curie-Weiss type temperature dependence of the positive AHE coefficient is repeatedly observed in our underdoped $Nd_{1-x}Sr_xNiO_2$ infinite-layer thin films (as shown in Fig. 3c) and is also common in recently reported nickelate superconductors [3,14,15,41]. It supports the existence of free local moments at high temperatures where the Kondo scattering is negligible and the resistivity is dominated by electron-phonon scattering. The negative values of the Weiss temperature $\Theta$ indicate AFM correlations in the localized-spin systems. Actually, a magnetic ground state is often obtained in theoretical studies [6,10,40,42,43]. A branch dispersion of magnetic excitations in undoped $NdNiO_2$ has recently been revealed using RIXS [44], suggesting a spin wave of strongly coupled, antiferromagnetically aligned spins on a square lattice. A recent NMR study shows the presence of AFM fluctuations and quasi-static AFM order in $Nd_{0.85}Sr_{0.15}NiO_2$ [45] thin film. However, the tendency towards a long-range AFM order is interrupted by the self-doping and Kondo screening effect, causing a paramagnetic state (Fig. 2a and Supplementary Fig. 6) [30] as reported so far for all $RNiO_2$ (R = rare-earth) parent materials [7,9].

Based on the present findings, Fig. 3d depicts a phase diagram of the $Nd_{1-x}Sr_xNiO_2$ showing a superconducting dome, combined with that in recent reports [3,15,46,47], and the characteristic temperature $T^*$, where the $R_H$ shows a maximum. As $x$ increases from the underdoped side, $T^*$ decreases monotonically and the $T^*$–$x$ curve separates the underdoped region into two parts: a Kondo scattering region and a metal. It is worth noting that the extension of the $T^*$–$x$ curve reaches the bottom of the superconducting dome (under field). Thus, instead of a simple Mott insulator, the Kondo physics must also play a crucial role in nickelate superconductors. Our results may indeed have some implications on the superconductivity. Based on the Kondo mechanism in the underdoped region, our phase diagram (see Fig. 3d) suggests that superconductivity emerges near the boundary that the Kondo effect is suppressed. As discussed previously [19], this may have important influence on the pairing symmetry of the superconductivity. The interplay of magnetic fluctuations and Kondo hybridization could potentially lead to $d + is$ pairing [19].

Putting together, the logarithmic temperature dependence of resistivity and $R_H$, the good agreement with the NRG, NCA, and Hamann predictions, the linear dependence of $\sigma_{xy}^{\mathrm{AHE}} \sim \sigma_{xx}$, and the linear temperature dependence of the dephasing rate, all support the presence of the magnetic Kondo scattering in the underdoped infinite-layer $Nd_{1-x}Sr_xNiO_2$ thin film. According to Zhang et al. [19], the presence of local moments cannot be ascribed to the Nd 4f moments since similar transport properties have also been observed in LaNiO$_2$ [9]. The first-principles band structure calculations [40] show that the Nd 5d orbitals in NdNiO$_2$ are hybridized with the Ni 3d orbitals, leading to small Fermi pockets of dominantly Nd 5d electrons in the Brillouin zone. Nd 5d conduction electrons have an electron density of n < 1 per Ni site, coupled to the localized Ni$^{1+}$ spin-1/2 of $3d_{x^2-y^2}$ orbital to form Kondo spin singlets, as in the Kondo lattice systems with a small concentration of conduction electrons [20]. The fitted magnetic impurity concentration (~0.03 per formula unit, see Supplementary Fig.5d) is comparable with that of dilute Kondo systems [39]. In conclusion, our experimental results strongly support the self-doped Mott-Kondo scenario for the underdoped Nd$_{1-x}$Sr$_x$NiO$_2$ infinite-layer thin films. The present findings shed new light on the underlying physics of the infinite-layer nickelates and a possibly novel mechanism of unconventional superconductivity. It would improve our understanding of the newly discovered superconductivity in nickelates.

This work was supported by the National Natural Science Foundation of China (Grants No. 92065110, No. 11974048, No. 12074334, No. 12174429, No. 12022407), the National Basic Research Program of China (Grants No. 2014CB920903 and No. 2013CB921701), and the Strategic Priority Research Program of the Chinese Academy of Sciences (Grant No. XDB33010100) for their financial support. This work was also supported by the High Magnetic Field Laboratory of Chinese Academy of Sciences at Hefei.

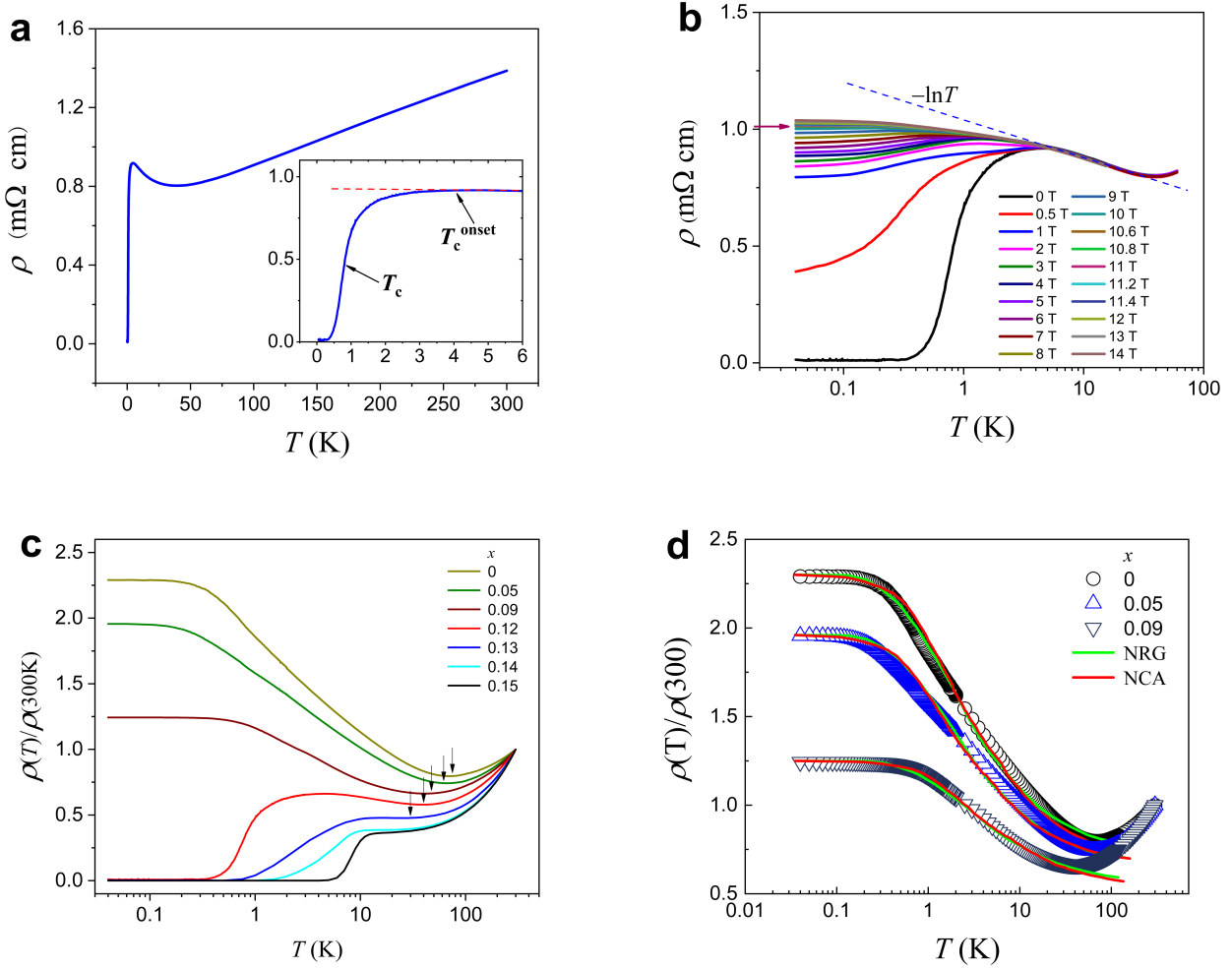

**Fig. 1| Temperature-dependent resistivity $\rho(T)$ for the underdoped Nd$_{1-x}$Sr$_x$NiO$_2$ thin films. a**, The temperature dependence of resistivity of the Nd$_{0.88}$Sr$_{0.12}$NiO$_2$ film at zero magnetic field. $T_c$ and $T_c^{onset}$, marked by black arrows, are 0.79 K and 4.06 K, respectively. The inset shows the determination of $T_c^{onset}$. **b**, Isomagnetic $\rho(T)$ curves of the Nd$_{0.88}$Sr$_{0.12}$NiO$_2$ film measured at different applied magnetic field $H$. **c**, The zero-field temperature dependence of resistivity of the underdoped Nd$_{1-x}$Sr$_x$NiO$_2$ films with a Sr doping level $x$ from 0.00 to 0.15. The arrows indicate the corresponding resistivity minima. **d**, The NRG (green) and NCA (red) fits are shown for the underdoped samples with x = 0, 0.05, and 0.09, and $T_K$ = 3.5 K, 3.5 K and 5.5 K, respectively (see details in Supplementary Fig. 4).

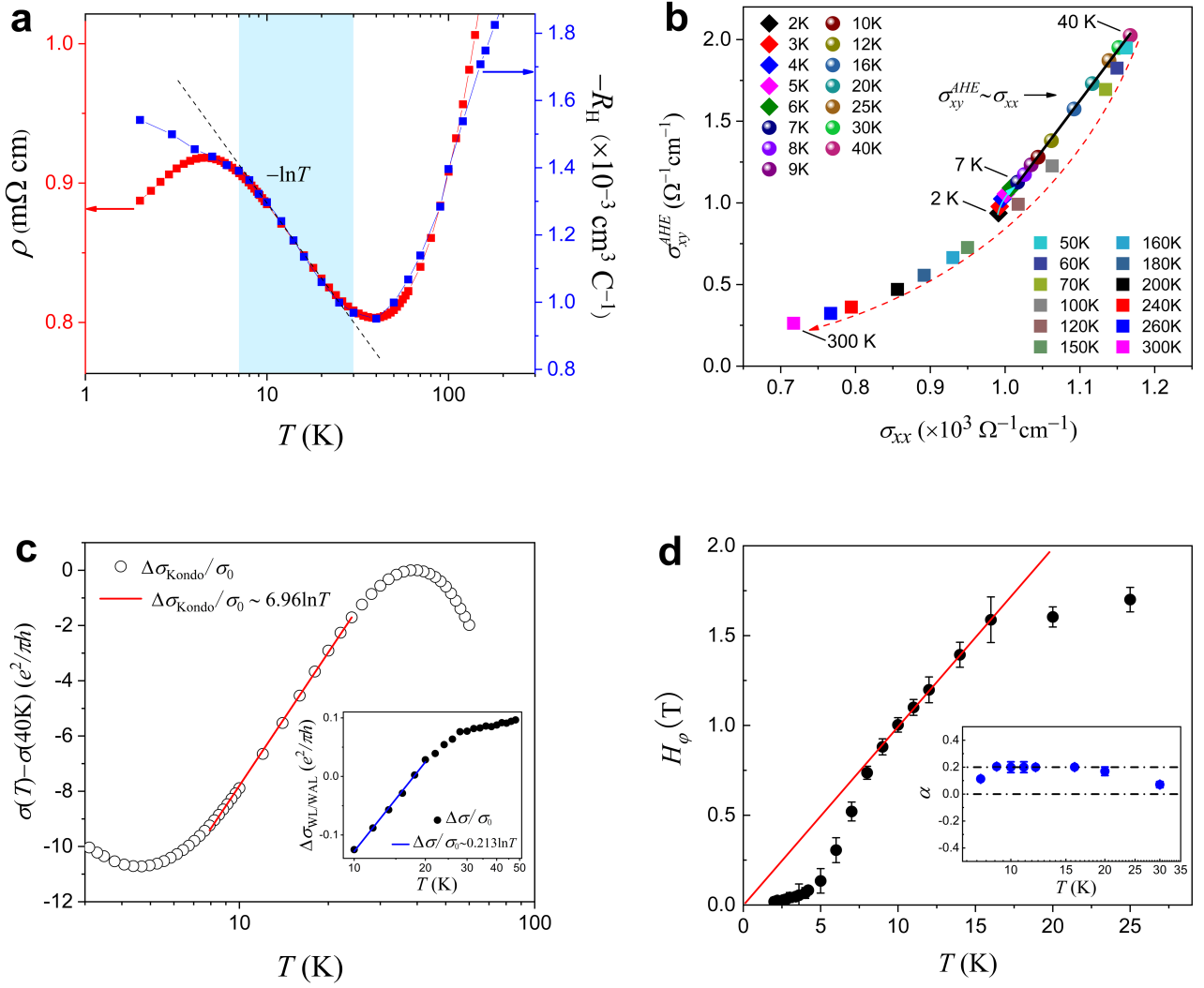

**Fig. 2 | Scaling behavior of anomalous Hall conductivity and conductance corrections for Nd$_{0.88}$Sr$_{0.12}$NiO$_2$.**
**a**, Logarithmic temperature dependence of the resistivity (red) and the Hall coefficient (blue) of the Nd$_{0.88}$Sr$_{0.12}$NiO$_2$ film. The light cyan area represents the Kondo region. **b**, Plot of AHE conductivity $\sigma_{xy}^{AHE}$ vs conductivity $\sigma_{xx}$ of the Nd$_{0.88}$Sr$_{0.12}$NiO$_2$ film over the entire temperature range. Since $\rho_{xy} \sim \rho_{xx}/10^3$, we can simplify the anomalous Hall conductivity as $\sigma_{xy}^{AHE} = -\rho_{xy}^{AHE}/\rho_{xx}^2$, and here $\rho_{xy}^{AHE} \equiv (R_H - R_0) \cdot H$. A linear dependence (solid black line) of $\sigma_{xy}^{AHE} \sim \sigma_{xx}$ is obvious in a temperature range of 7–40 K. **c**, Zero field conductance correction for the Nd$_{0.88}$Sr$_{0.12}$NiO$_2$ film due to the Kondo effect, i.e., $\Delta\sigma_{Kondo} = \sigma(T) - \sigma(T_{min} = 40K)$. Data are extracted from Fig. 1b and the solid red line is the $lnT$ fits. Inset: Conduction correction due to the WL/WAL effect, which is obtained with $\Delta\sigma_{WL/WAL} = \sigma(T)|_{H=0} - \sigma(T)|_{H=4T}$ (solid circles). The solid blue line is the $lnT$ fit for obtaining $\Delta\kappa = \alpha p$, and here $\sigma_0 = e^2/\pi h$. **d**, Dephasing field $H_\varphi$ versus temperature for the Nd$_{0.88}$Sr$_{0.12}$NiO$_2$ film. The straight red line is the linear fit. The inset shows the corresponding $\alpha$ values, which are nearly constant in the temperature range of 9–20 K.

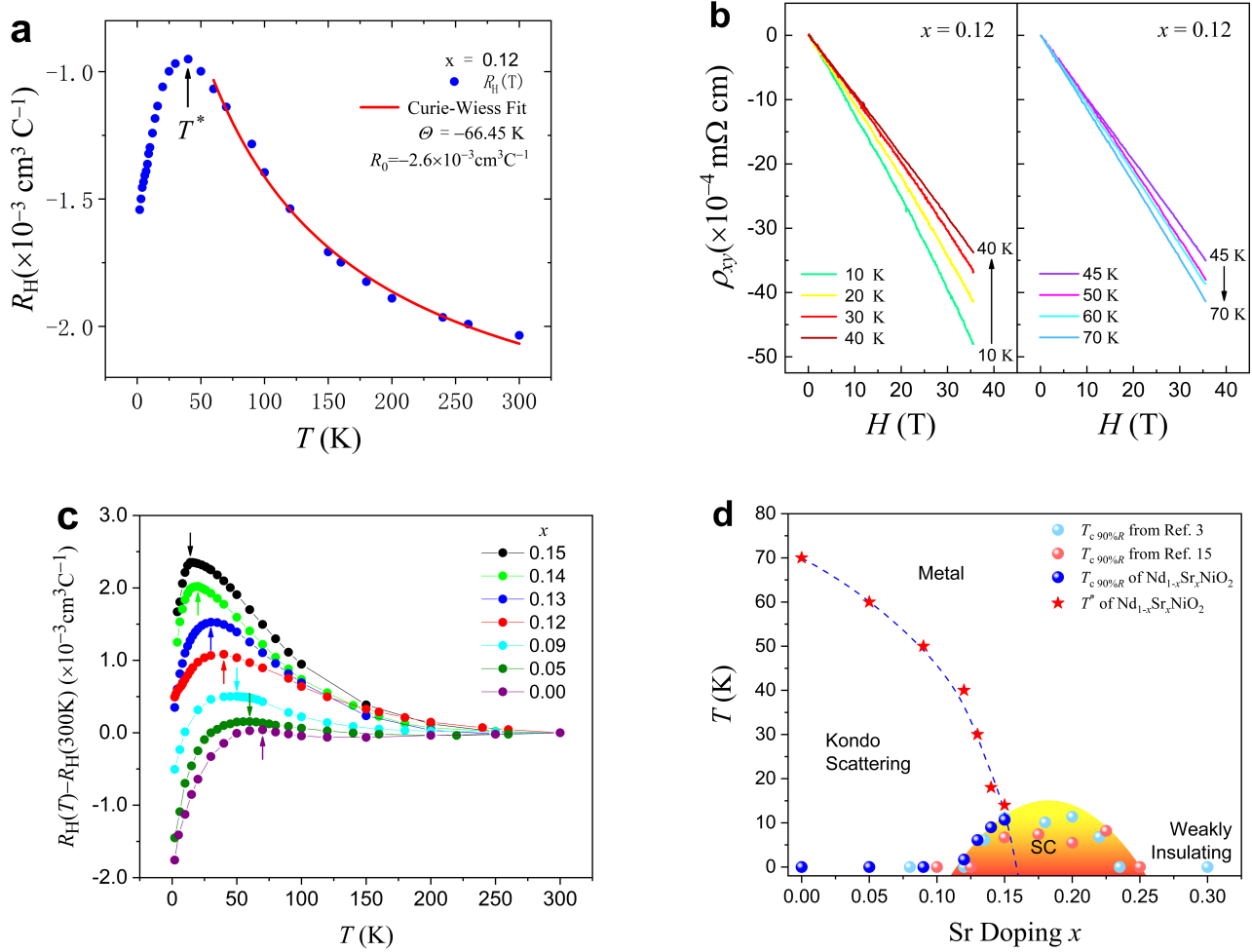

**Fig. 3| Kondo scattering dominated region in phase diagram of Nd$_{1-x}$Sr$_x$NiO$_2$. a,** Hall coefficient $R_H(T)$ acquired in a field of 9 T for the Nd$_{0.88}$Sr$_{0.12}$NiO$_2$ film with a maximum around 40 K. **b,** Measured Hall resistivity, $\rho_{xy}$, versus magnetic field, $H$, at different temperatures. The Hall resistivity $\rho_{xy}$ shows no obvious deviation from linear dependence on magnetic field $H$ up to 35 T at all temperatures. **c,** Hall coefficient $R_H(T) - R_H(300K)$ acquired in a field of 9 T for the underdoped Nd$_{1-x}$Sr$_x$NiO$_2$ films, with a maximum around $T^*$, at which the resistivity normally shows a minimum (see the arrows in Fig.1c). The corresponding characteristic temperatures $T^*$ are indicated by the arrows. **d,** The superconducting transition temperature $T_c$ (circles) and the characteristic temperature $T^*$ (stars) where the $R_H$ shows a maximum (Fig. 3c). The $T_{c90\%R}$ is defined to be the temperature at which the resistivity drops to 90% of the value at the onset of the superconductivity. The cyan and orange circles represent the average $T_{c90\%R}$ adapted from references [3,15]. The blue circles represent the $T_{c90\%R}$ of the samples shown in this study.

# Supplementary Material
# Kondo scattering in underdoped Nd$_{1-x}$Sr$_x$NiO$_2$ infinite-layer superconducting thin films

T. N. Shao[1], Z. T. Zhang[1], Y. J. Qiao[1], Q. Zhao[1], H. W. Liu[1*], X. X. Chen[1], W. M. Jiang[1], C. L. Yao[1], X. Y. Chen[1], M. H. Chen[1], R. F. Dou[1], C. M. Xiong[1], G. M. Zhang[2,3*], Y.-F. Yang[4,5,6*], J. C. Nie[1*]

[1]*Department of Physics, Beijing Normal University, Beijing 100875, China*

[2]*State Key Laboratory of Low-Dimensional Quantum Physics and Department of Physics, Tsinghua University, Beijing 100084, China*

[3]*Frontier Science Center for Quantum Information, Beijing 100084, China*

[4]*Beijing National Laboratory for Condensed Matter Physics and Institute of Physics, Chinese Academy of Sciences, Beijing 100190, China*

[5]*School of Physical Sciences, University of Chinese Academy of Sciences, Beijing 100190, China*

[6]*Songshan Lake Materials Laboratory, Dongguan, Guangdong 523808, China*

*Corresponding author. E-mail: jcnie@bnu.edu.cn (J.C.N.); yifeng@iphy.ac.cn (Y.-F.Y.); gmzhang@tsinghua.edu.cn(G.M.Z.); haiwen.liu@bnu.edu.cn (H.W.L.)

## Methods

### Sample preparation and characterization

The La$_{1-x}$Sr$_x$NiO$_3$ target was prepared by a stoichiometric mixture of SrCO$_3$, Nd$_2$O$_3$, and NiO powders used a solid-state reaction in air at 1350°C for 12 hours. The products were ground and reheated of this reaction, and this process was repeated three times. The powder was pressed into 2.5cm sheets and then calcined for 24 h at 1200°C in air. The heating and cooling of the sintering were kept at a rate of 10 °C/min. Precursor perovskite thin films were grown on the TiO$_2$-terminated SrTiO$_3$ substrate (NH$_4$F buffered HF etching solution) by pulsed laser deposition from polycrystalline target ablated with a KrF excimer laser (wavelength = 248 nm). During the deposition, the oxygen partial pressure was 150 mTorr and the substrate temperature was at 600°C. After deposition, the films were kept at 600°C for 10 minute and cooled to room temperature at a rate of 10°C per minute in the same oxygen partial pressure. The laser energy was 1.2 J/cm$^2$ and the frequency was 4 Hz. For the very thin as-grown films less than 10 nm, nominal thickness was usually determined by the number of laser shots and the calibrated growth rate [48]. The soft-chemistry reduction method [1,3,7,41,49,50] was used to acquire the infinite-layer nickelate phase. The as-grown precursor perovskite thin films were placed in Pyrex glass tubes with calcium hydride (CaH$_2$) powder (∼ 0.1 g) after loosely wrapping them with aluminum foil. The glass tubes were pumped to a vacuum about 1×10$^{-5}$ Torr. Then it was heated up to 320°C and keep for 2 hours, with the heating and cooling rates of 10 °C/min.

The structural properties of the films were measured by using a x-ray diffractometer (Shimadzu XRD-6000). The valence state and the atomic ratio of the relevant elements of the reduced sample was investigated by a Thermo Scientific ESCSLAB 250Xi XPS. For very thin as-grown films, the mean grain size (out-of-plane) $d$ is well consistent with the film thickness. The full width at half maximum (FWHM) of the most intense XRD peaks, (002) of the perovskite Nd$_{1-x}$Sr$_x$NiO$_3$ and (002) of the infinite-layer Nd$_{1-x}$Sr$_x$NiO$_{2,}$ are routinely used to determine d using the Scherrer formula, $d=(0.9\lambda/\beta_c\cos\theta)$, where $\lambda$ is the wavelength of the filament used in the XRD machine, $\beta_c$ is the FWHM, $\theta$ is the angle of the same peak.

## Electrical Measurements

The normal state electrical-transport and magneto-transport measurements were carried out using the Van der Pauw configuration in a physical property measurement system (PPMS, Quantum Design). Electrode contacts to the samples were bonded by ultrasonic wire bonding (Al wire of 25 μm diameter). The DC current for the normal state measurements was 10 μA. The sub-Kelvin measurements were performed with a lock-in amplifier (Stanford Research System, SR830) in AC mode in a dilution refrigerator Triton-400 system (Oxford Instruments, base temperature < 10 mK). Standard four probe resistance measurements were made with sufficiently low excitation current (50 nA at 7.9 Hz) to avoid any heating of the electrons at the lowest temperature. The high magnetic field (0-35 T) measurements were carried out by the Hefei Steady-state High Magnetic Field Facilities in High Magnetic Field Laboratory, Chinese Academy of Sciences and University of Science and Technology of China.

## Results and Discussion

Supplementary Fig. 1a illustrates the structural properties of the representative $Nd_{1-x}Sr_xNiO_3$ thin films before the topotactic reduction, characterized by X-ray diffraction (XRD) measurements. The thickness of the as-grown film is about 8.48 nm. After chemical reduction, the film peaks appear at 26.70° and 55.06°, which are the characteristic peaks of the infinite layer $Nd_{1-x}Sr_xNiO_2$, corresponding to (001) and (002) reflections, respectively. The corresponding c-axis lattice constant is about 3.333 Å, indicating a slight expansion of the c-axis lattice constant comparing with that (3.28 Å) of the undoped $NdNiO_2$ film [15]. According to the previous reports [3,15], in our case, the empirical Sr doping level x is about 0. 12, consistent with that of the stoichiometric target.

In order to further quantitatively analyze the chemical composition and elemental valence of the film, X-ray photoelectron spectroscopy (XPS) measurements were carried out. Supplementary Fig. 1b and Fig. 1c show the XPS spectra of Ni $2p_{3/2}$ and Nd $3d_{5/2}$ of the representative $Nd_{1-x}Sr_xNiO_2$ sample treated by the topotactic hydrogen reduction. The peak at the binding energy of 852.5 eV corresponds to $Ni^+$ cations, and the left side peak at the binding energies 854.6 eV originates from $Ni^{2+}$ cations [51]. A quantitative analysis, as shown in Supplementary Fig. 1b, gives the relative concentrations of $Ni^+$ and $Ni^{2+}$ cations. It turns out that the percentages of $Ni^+$ and $Ni^{2+}$ cations are 71.8% and 28.2%, respectively. The high resolution XPS spectrum of Nd $3d_{5/2}$ core level line is displayed in Supplementary Fig. 1c. The main peak of Nd $3d_{5/2}$ at the binding energy of 982.2 eV corresponds to $Nd^{3+}$ cations, and the one at 978.2 eV is a satellite peak of the main one, while the shoulder at low binding energy (972.7 eV) is expected to contain a significant contribution of the $KL_{23}L_{23}$ Auger of oxygen which is consistently observed in most oxides, such as $Nd_2NiO_4$ [52]. The calculated atomic ratio of Nd:Ni is about 0.884:1, which is well consistent with the result of the XRD data. Both the XRD and XPS results indicate that the composition of the representative infinite-layer thin film is $Nd_{0.88}Sr_{0.12}NiO_2$, which is in the underdoped region. In addition, a nickel oxidation state of +1.28 may indicate a little excess oxygen content in the infinite-layer thin film.

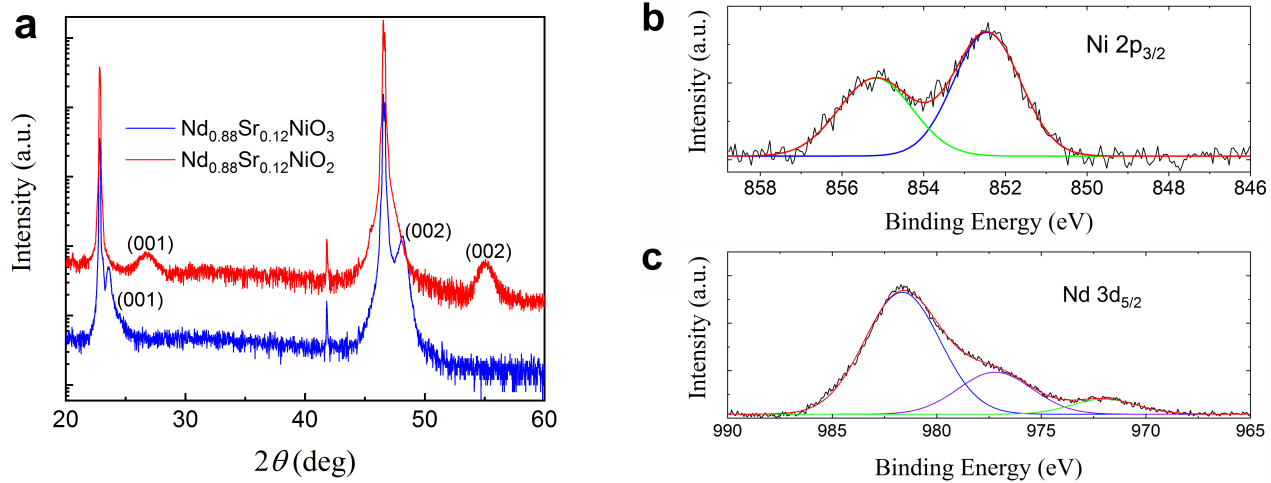

**Supplementary Fig. 1 | Characterization of the representative nickelate thin film.** (**a**) Structural characterization of the nickelate film on SrTiO$_3$ substrate. The XRD $\theta$–$2\theta$ scans of the representative Nd$_{0.88}$Sr$_{0.12}$NiO$_3$ film before and after the topotactic reduction at 320°C for 2 hours. XPS spectra of Ni 2p (**b**) and Nd 3d (**c**) of the representative Nd$_{0.88}$Sr$_{0.12}$NiO$_2$ sample treated by the topotactic hydrogen reduction.

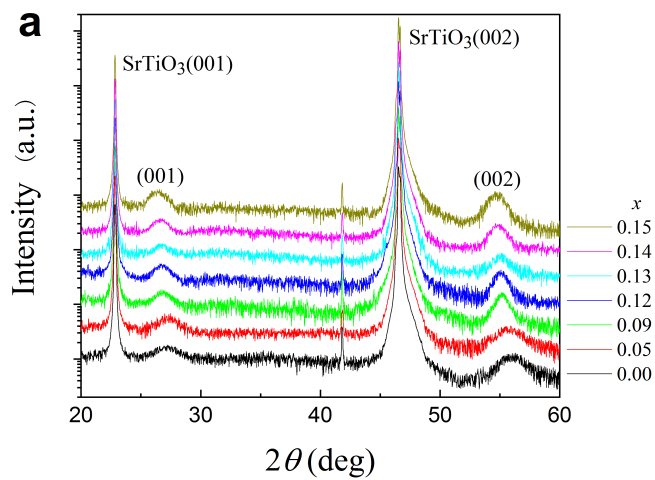 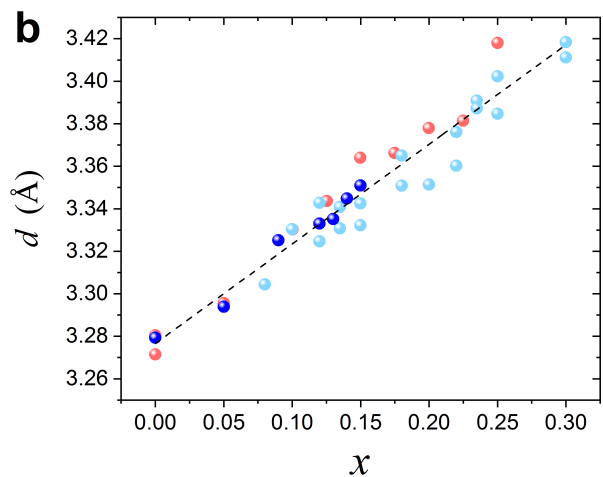

**Supplementary Fig. 2| Structural characterization of the underdoped Nd$_{1-x}$Sr$_x$NiO$_2$ infinite layer thin films with different Sr doping.** (**a**) The XRD patterns of the underdoped Nd$_{1-x}$Sr$_x$NiO$_2$ infinite layer thin films. The intensity is vertically displaced for clarity. (**b**) Room-temperature *c*-axis lattice constants *d* as a function of Sr doping level *x*. The blue circles represent the data extracted from the $\theta$-$2\theta$ scans in (**a**) and the cyan and orange circles represent the data adapted from references [3,15]. The dashed line is a guide to the eye.

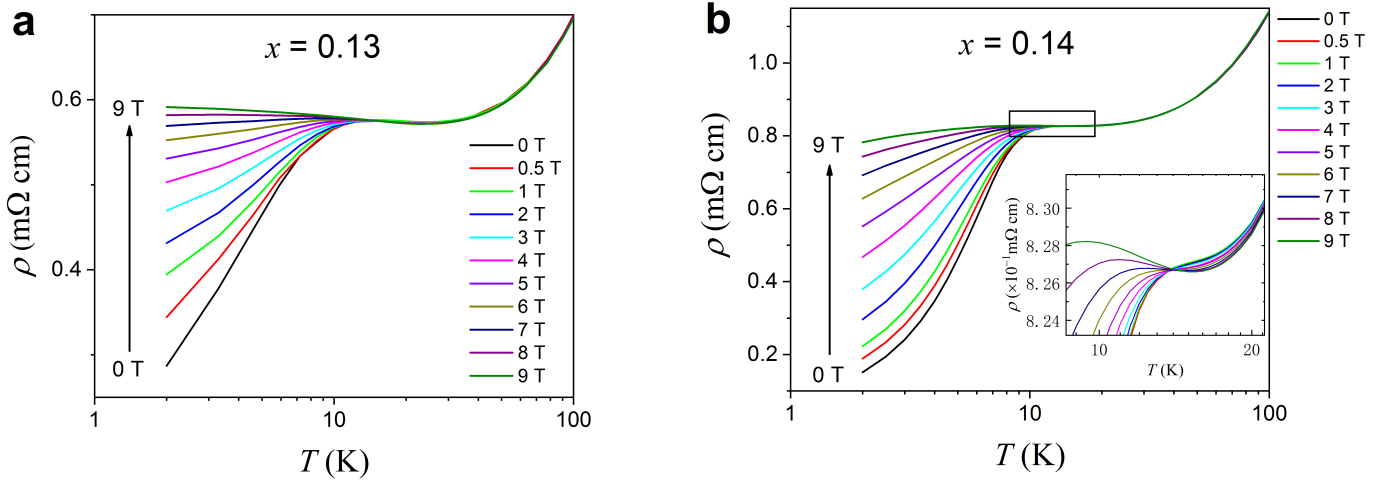

**Supplementary Fig. 3| Isomagnetic ρ(T) curves of the other under-doped $Nd_{1-x}Sr_xNiO_2$ thin films with a Sr doping level x = 0.13 (a) and 0.14 (b).** For the sample with x = 0.13 (**a**), in all magnetic fields, the resistivity-temperature curve shows a low-temperature minimum, followed by an upturn that is proportional to *lnT* with decrease of temperature. (**b**) For the sample with x = 0.14, in a strong magnetic field (e.g., *H > 7.0* T), the resistivity-temperature curve shows a low-temperature minimum followed by a small upturn that is proportional to *lnT* with decrease of temperature (see the inset). The inset is an enlarged view of the block in Supplementary Fig. 3b.
5

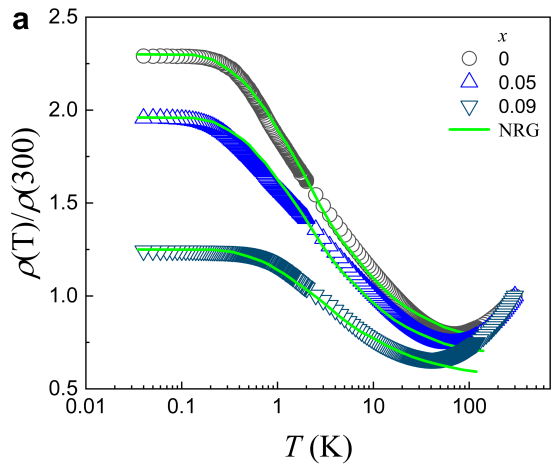 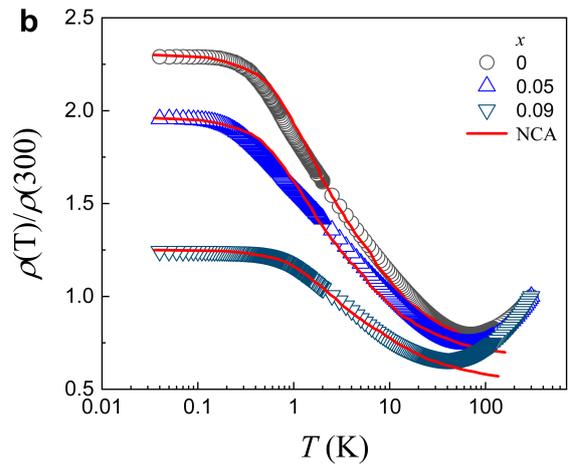

**Supplementary Fig. 4| Temperature dependence of resistivity of the underdoped Nd$_{1-x}$Sr$_x$NiO$_2$ film, fitted by NRG (a) and NCA (b) in temperature regime from 0.04 K to 100 K.** The experimental data show the resistivity as a function of T from 0.04 K to 300 K. The fitting curves give T$_K$ = 3.5 K, 3.5 K and 5.5 K for x = 0, 0.05 and 0.09 samples, respectively. The NCA fitting is compared with Fig.9.29 in Hewson's book [20]. The NRG fitting is compared with Fig.13 in the work by Costi et al. [20].

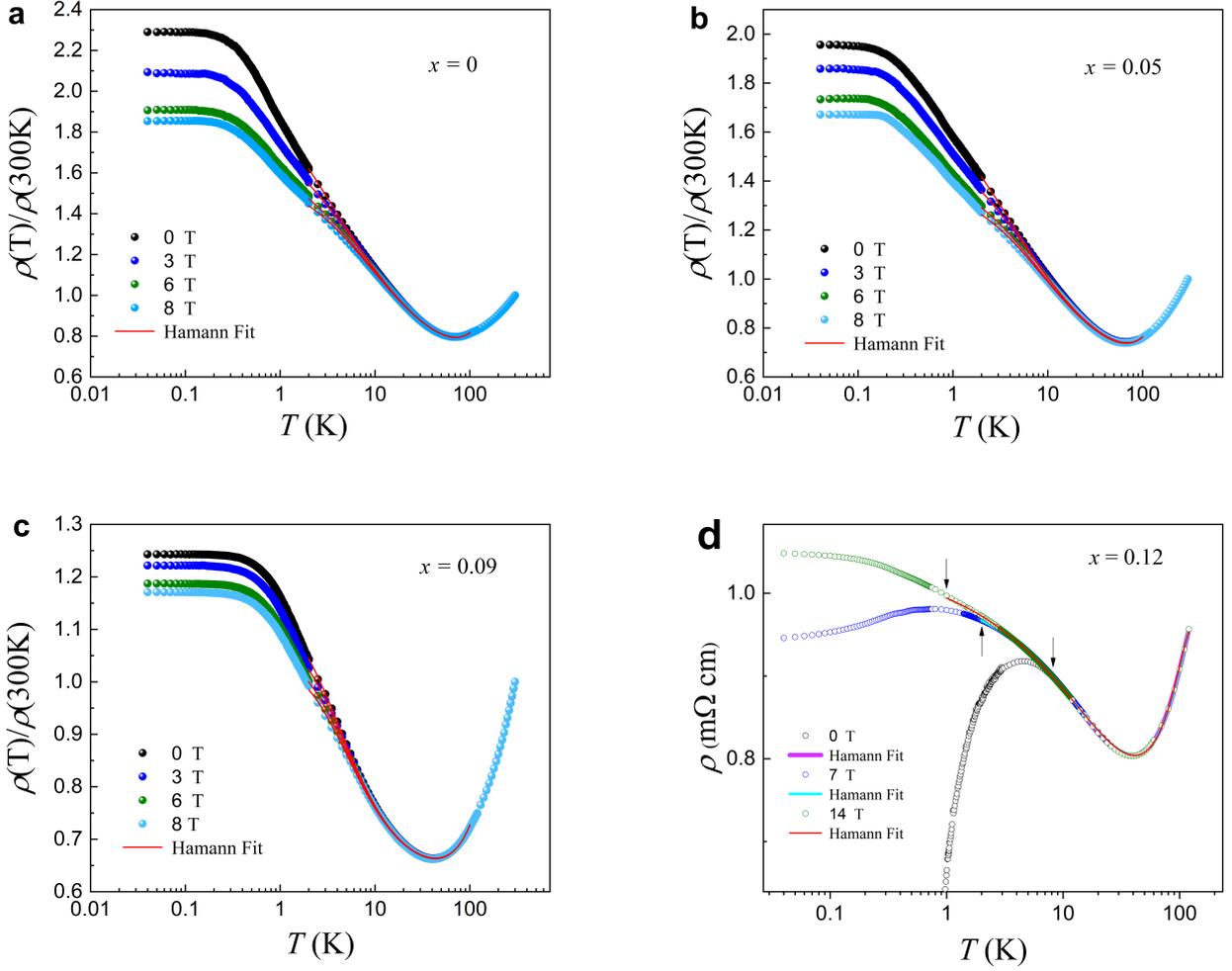

**Supplementary Fig. 5| Temperature dependence of resistivity of the underdoped Nd$_{1-x}$Sr$_x$NiO$_2$ film, fitted by Hamann Model. (a)**, **(b)** and **(c)** show the resistivity as a function of T from 0.04 K to 300 K for x = 0, 0.05 and 0.09 samples, respectively. The data without magnetic field are fitted by the Hamman model in temperature regime from 2 K to 100 K (Eq. 1 and Eq.2 in the main text), with the Kondo temperature $T_K$ = 3.5 K, 4.0 K and 4.38 K and s = 0.68, 0.59 and 0.23 for x = 0, 0.05 and 0.09 samples, respectively. The data under magnetic field are fitted by the modified Hamman model [31]. Under magnetic field, the magnetoresistance can be obtained by the: $\rho_K(T/T_K) = \rho_0 \left\{1 - \frac{\ln(T/T_K)}{[\ln^2(T/T_K)+s(s+1)\pi^2]^{-1/2}}\right\} \cdot \left[1 - F^2\left(\frac{g\mu_B H}{k_B(T+T_K)}\right)\right]$. Here, F(x) is the Brillouin function for spin *s* and *g* denotes the Landé *g*-factor. Related discussions are shown in Supplementary Note 2. **(d)** shows the resistivity of the x = 0.12 sample as a function of T from 0.04 K to 120 K. Solid lines are fitting result using modified Hamman model, yielding the Kondo temperature $T_K$ of about 15.6 K, the effective spin s = 0.225, and the magnetic impurity concentration *c* of about 5.83×10$^{20}$cm$^{-3}$, i.e., 0.03 magnetic-ion per unit cell in Nd$_{0.88}$Sr$_{0.12}$NiO$_2$, which is reasonable and comparable with that of dilute Kondo systems [29, 30].

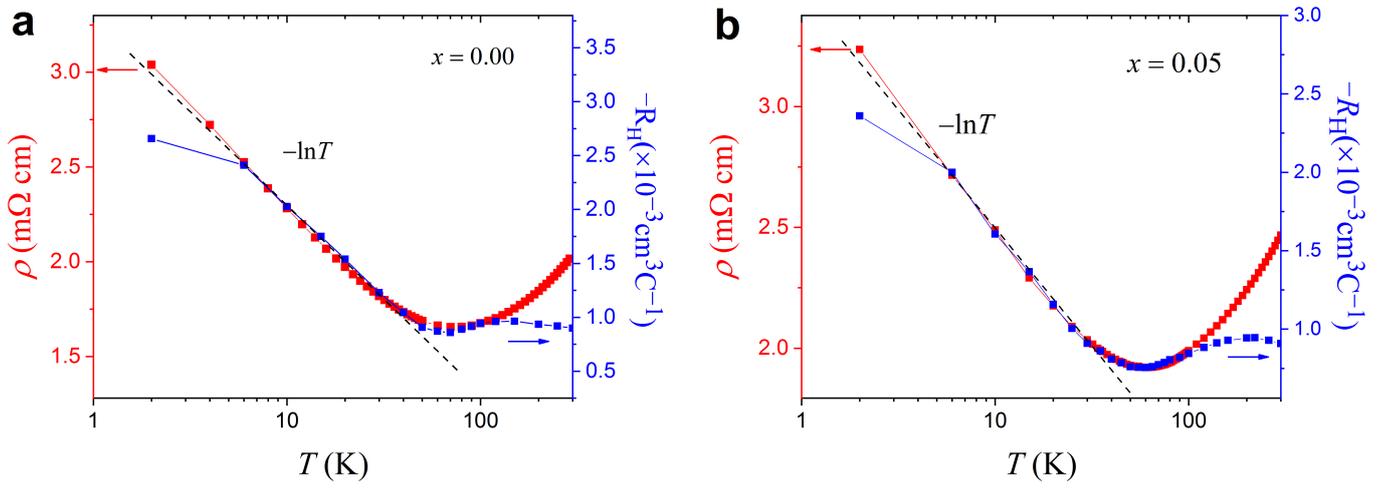

**Supplementary Fig. 6| Logarithmic temperature dependence of the zero-field resistivity and the Hall coefficient of the other under-doped Nd$_{1-x}$Sr$_x$NiO$_2$ thin films with (a) x=0.00 and (b) x=0.05.** For the underdoped nickelates, the $R_H$ resembles the $\ln T$ dependence of the resistivity for temperatures in a wide temperature range. The deviation of $R_H(T)$ from the $\ln T$ dependence at the lowest temperatures is obvious.

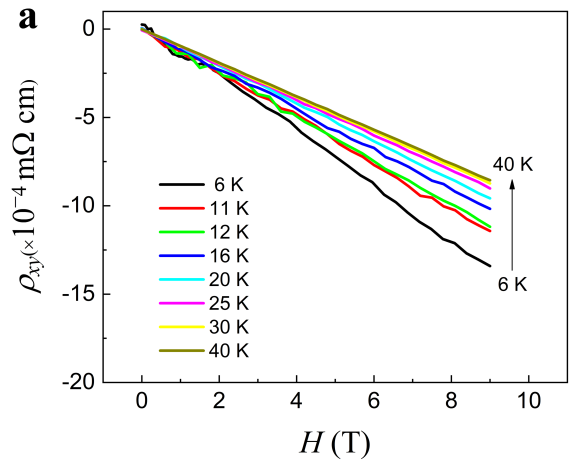 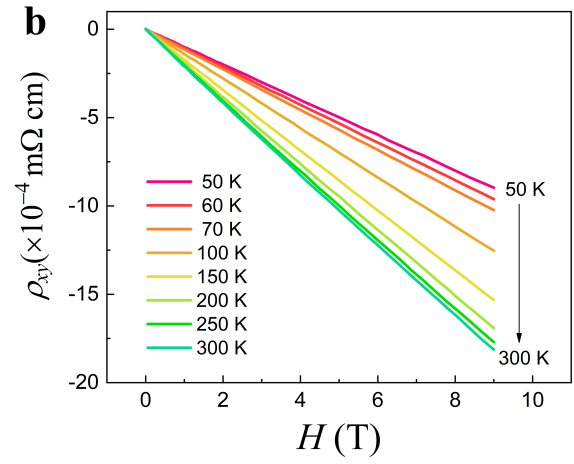

**Supplementary Fig. 7| Field dependence of Hall resistivity of the representative Nd$_{0.88}$Sr$_{0.12}$NiO$_2$ film.** Measured Hall resistivity, $\rho_{xy}$, versus magnetic field, $H$, at different temperatures: (**a**) 6–40 K and (**b**) 50–300 K. The Hall resistivity $\rho_{xy}$ shows no obvious deviation from linear dependence on magnetic field H up to 9 T at all temperatures.

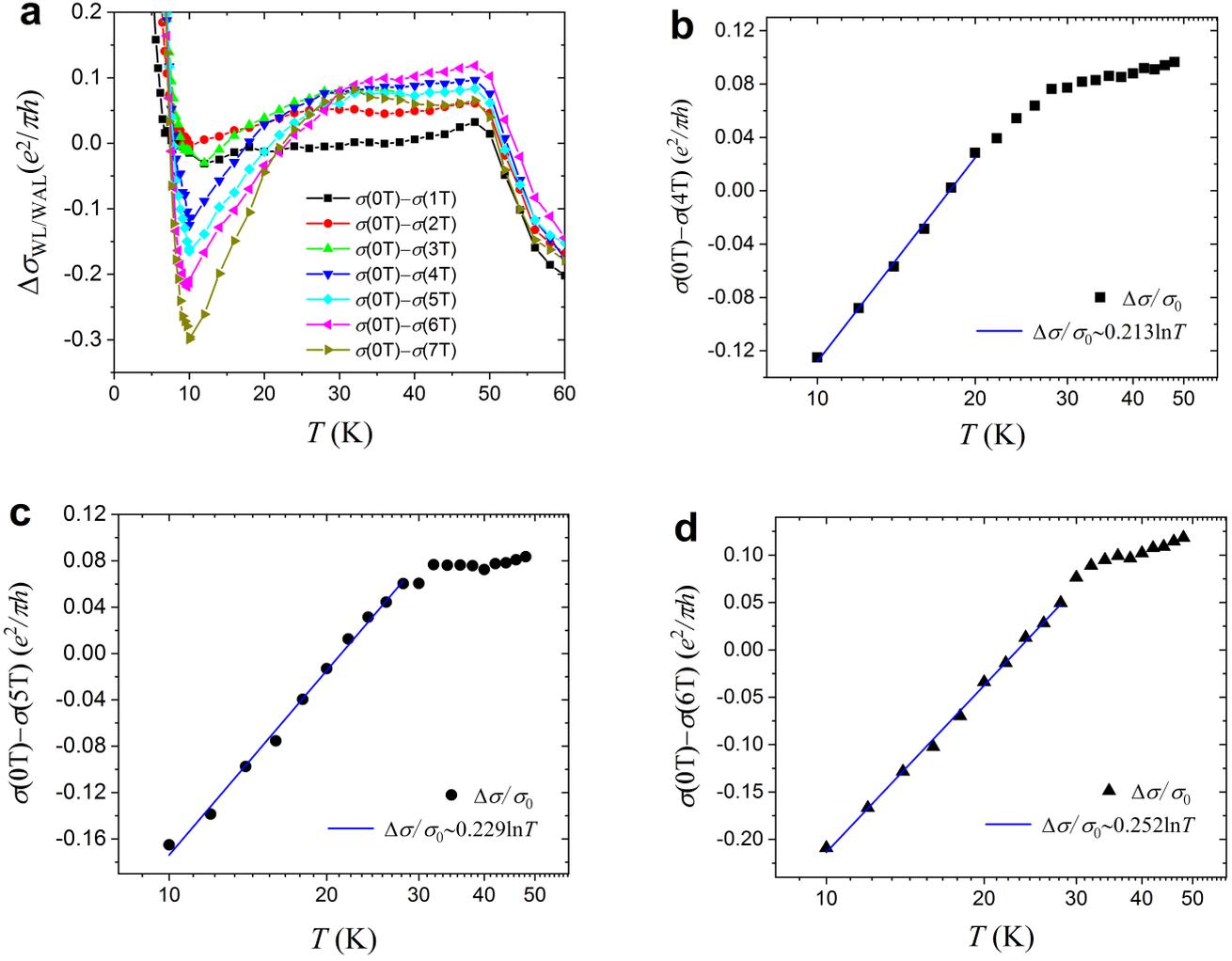

**Supplementary Fig. 8| WAL corrections of conductivity of the representative $Nd_{0.88}Sr_{0.12}NiO_2$ film.** The WL/WAL effect is insignificant at higher magnetic fields, the conductivity correction $\Delta\sigma_{WL/WAL} = \sigma(T)|_{H=0} - \sigma(T)|_{H=4T}$ represents the WL/WAL correction [53]. (**a**) Conduction corrections due to the WAL effect, which are obtained with $\Delta\sigma_{WL/WAL} = \sigma(T)|_{H=0} - \sigma(T)|_{H\neq 0}$. Logarithmic temperature dependence of conduction correction in a range of 10–48 K, with $\Delta\sigma_{WL/WAL} = \sigma(T)|_{H=0} - \sigma(T)|_{H=4T}$. (**b**), $\Delta\sigma_{WL/WAL} = \sigma(T)|_{H=0} - \sigma(T)|_{H=5T}$ (**c**), and $\Delta\sigma_{WL/WAL} = \sigma(T)|_{H=0} - \sigma(T)|_{H=6T}$ (**d**), respectively. The solid lines are the lnT fits for obtaining $\Delta\kappa = \alpha p$. As shown in Supplementary Fig. 7, $\alpha \approx 0.20$ can be maintained for a wide temperature range of 10–20 K. Thus, one can also obtain $H_\varphi$ have a linear power-law dependence satisfying $H_\varphi \sim T^p$ with $p \approx 1$, which is consistent with the analysis of HLN formula in Fig.2d.

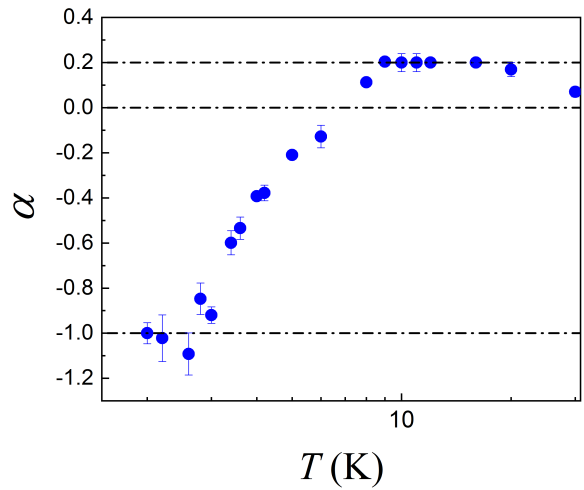

**Supplementary Fig. 9 The $\alpha$ values of the HLN fitting for the representative Nd$_{0.88}$Sr$_{0.12}$NiO$_2$ film.** A small value of $\alpha \approx 0.20$ can be maintained for a wide temperature range of 10–20 K. Here, α is an effective constant in the modified Hikami-Larkin-Nagaoka formula, which depends on the relative strengths of magnetic scattering and spin-orbital coupling. Typically, e.g., *α = 1* for weak spin-orbit and magnetic scattering, *α=−1/2* for strong SOC and weak magnetic scattering, and *α=0* for strong magnetic scattering.

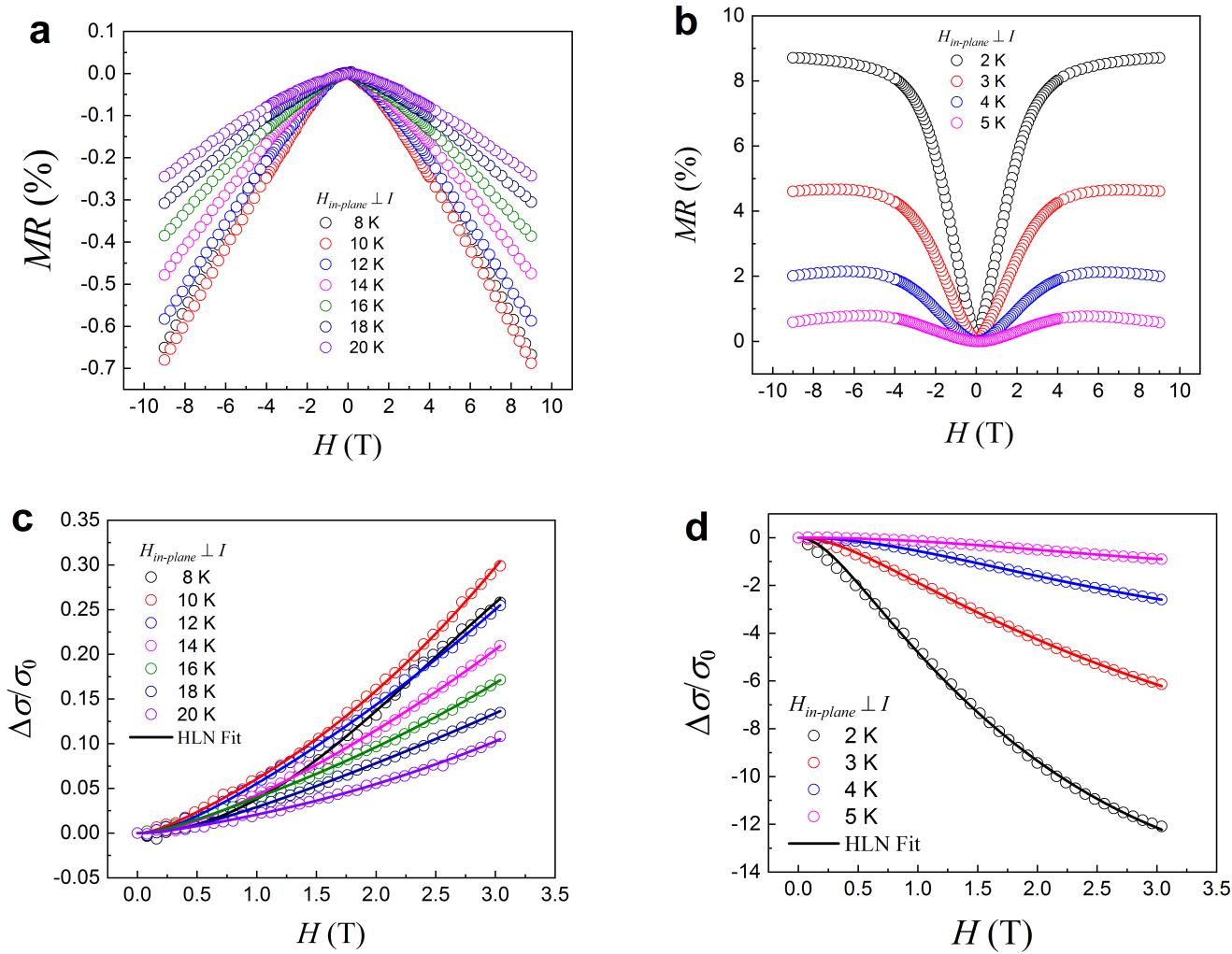

**Supplementary Fig. 10| In-plane transverse MR$_{//}$ of the representative Nd$_{0.88}$Sr$_{0.12}$NiO$_2$ film, fitted by HLN.** MR$_{//}$ (H // ab, H⊥I), at temperatures of 8–20 K (**a**) and 2–5 K (**b**). A sign change in MR$_{//}$ is seen below 7 K. (**c**) $\Delta\sigma/\sigma_0$ as a function of H at high temperatures. (**d**) $\Delta\sigma/\sigma_0$ as a function of *H* at low temperatures. The solid curves are fitting to the HLN Eq. (3) in the main text, by adding a $H^2$ term.

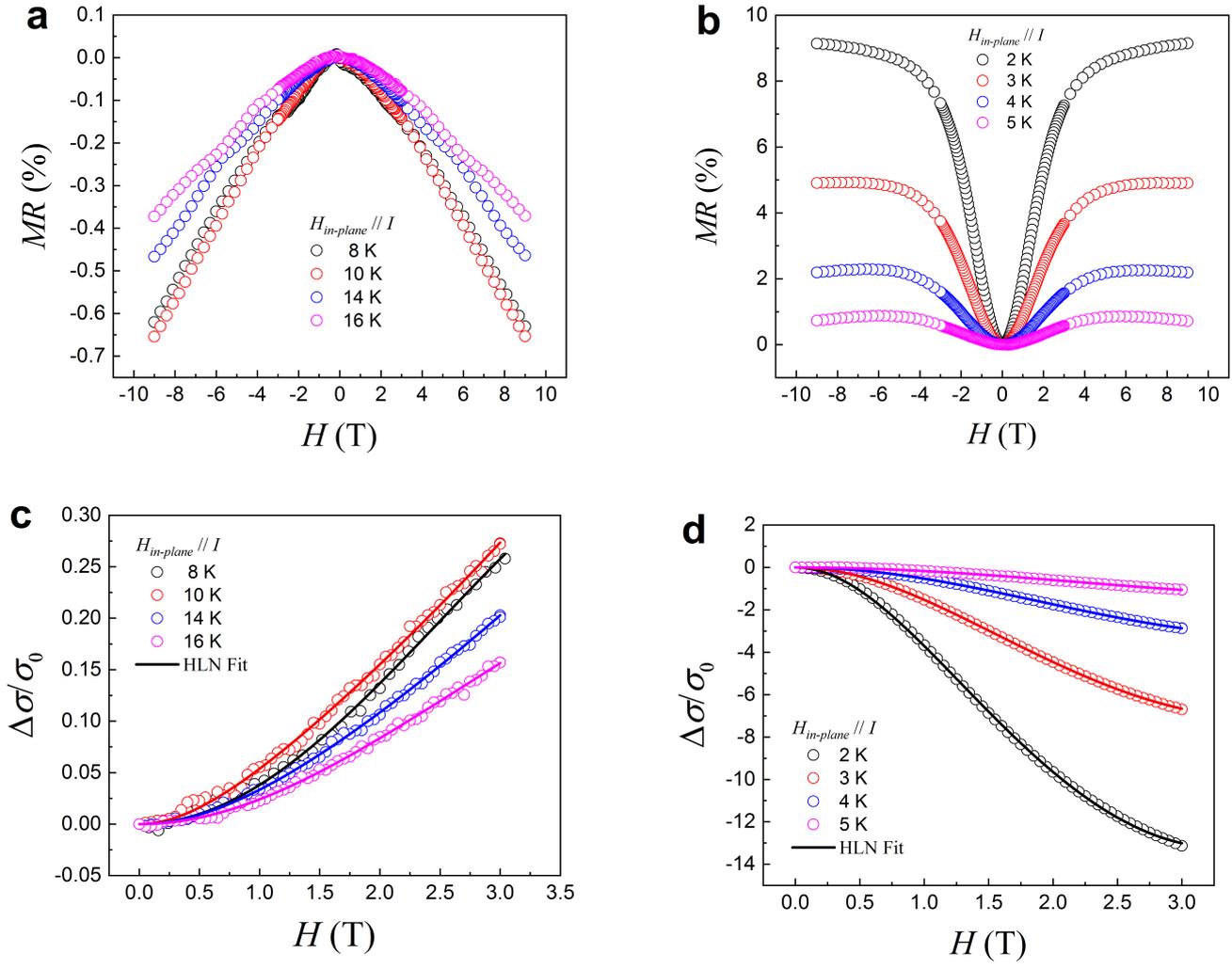

**Supplementary Fig. 11| In-plane longitudinal MR'$_{//}$ of the representative Nd$_{0.88}$Sr$_{0.12}$NiO$_2$ film, fitted by HLN.** In-plane longitudinal MR'$_{//}$ ($H$//ab, $H$//$I$), at temperatures of 8–16 K (**a**) and 2–6 K (**b**). A sign change in MR'$_{//}$ is seen below 7 K. (**c**) $\Delta\sigma/\sigma_0$ as a function of $H$ at high temperatures. (**d**) $\Delta\sigma/\sigma_0$ as a function of $H$ at low temperatures. The solid curves are fitting to the HLN Eq. (3) in the main text, by adding a $H^2$ term.

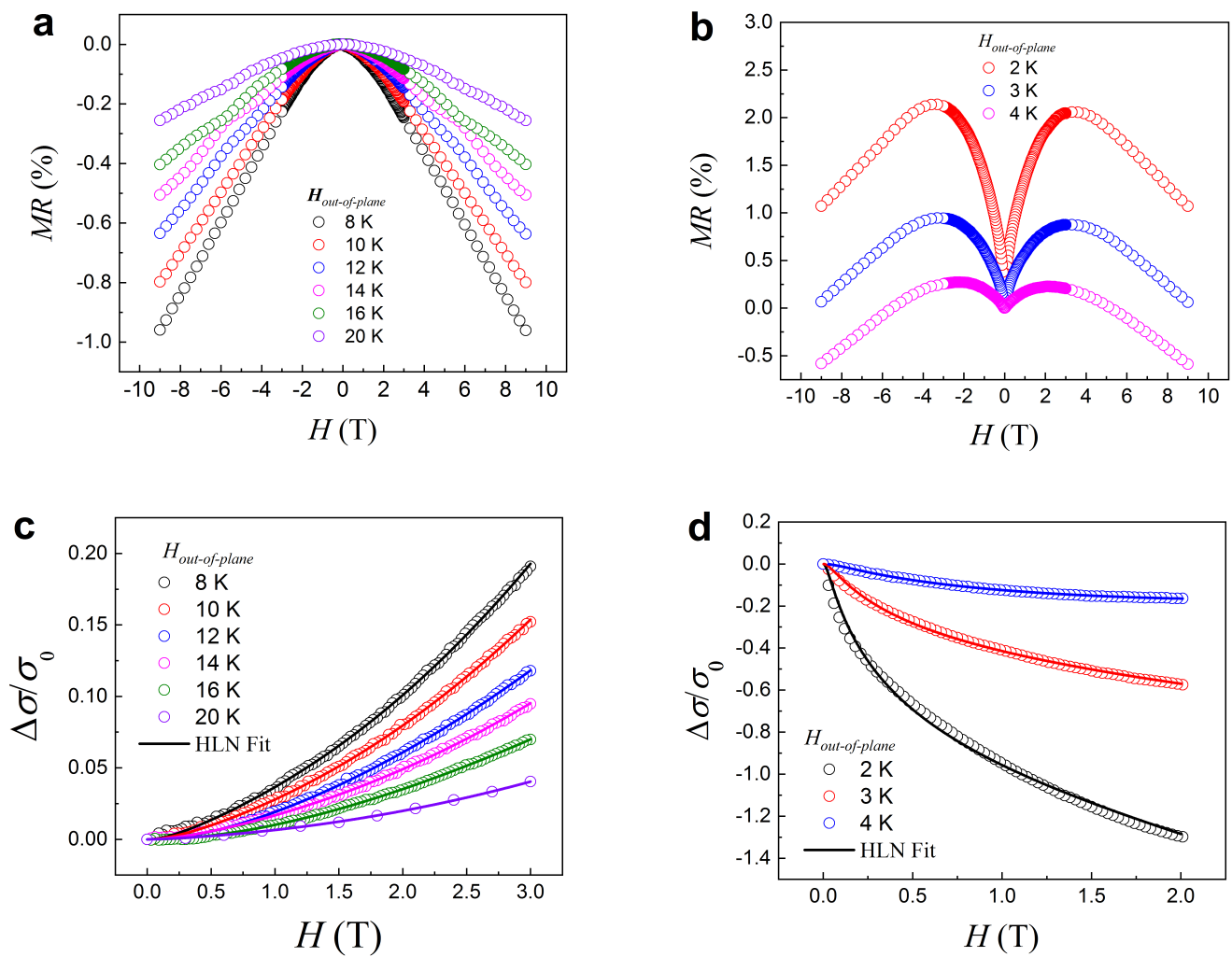

**Supplementary Fig. 12| Out-of-plane MR⊥ of the representative Nd$_{0.88}$Sr$_{0.12}$NiO$_2$ film, fitted by HLN.** MR⊥ ($H$⊥ab) at temperatures of 8–20 K (**a**) and 2–4 K (**b**). A sign change in MR⊥ is seen below 7 K. (**c**) $\Delta\sigma/\sigma_0$ as a function of H at high temperatures. (**d**) $\Delta\sigma/\sigma_0$ as a function of $H$ at low temperatures. The solid curves are fitting to the HLN Eq. (3) in the main text, by adding a $H^2$ term.

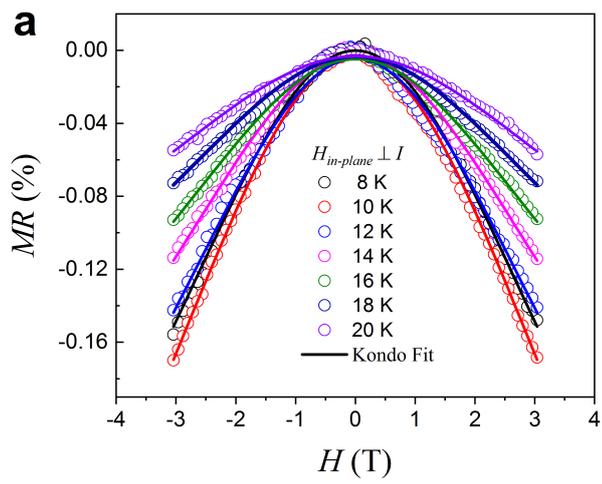 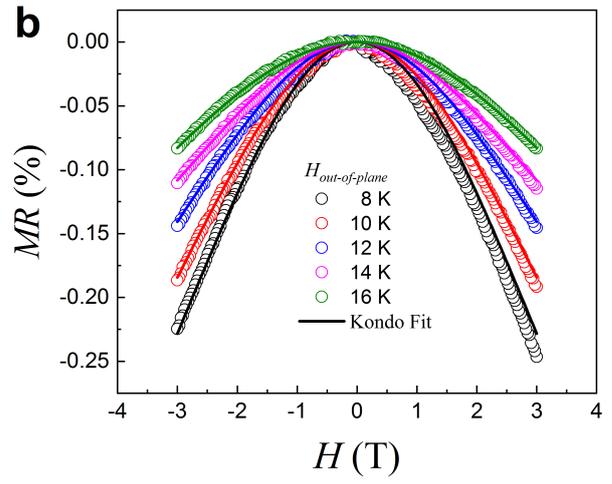

**Supplementary Fig. 13| Negative in-plane transverse MR$_{//}$ and out-of-plane MR$_\perp$ of the representative Nd$_{0.88}$Sr$_{0.12}$NiO$_2$ film, fitted by the Kondo model.** MR$_{//}$ ($H$ // ab, $H\perp I$) at temperatures of 8−20 K (a) and MR$_\perp$ ($H\perp$ab) at temperatures of 8−16 K (b). The solid curves are fitting to the Kondo model.

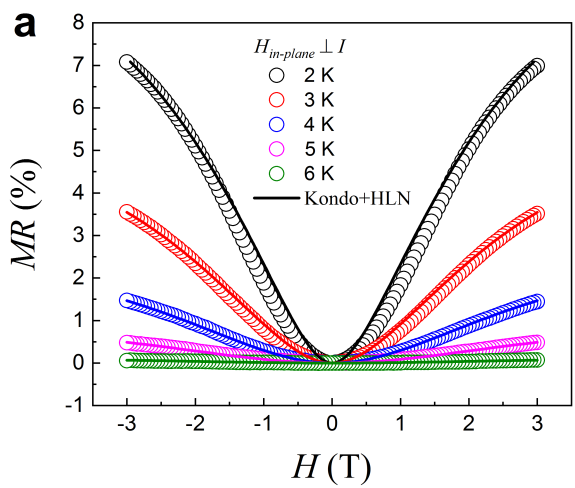 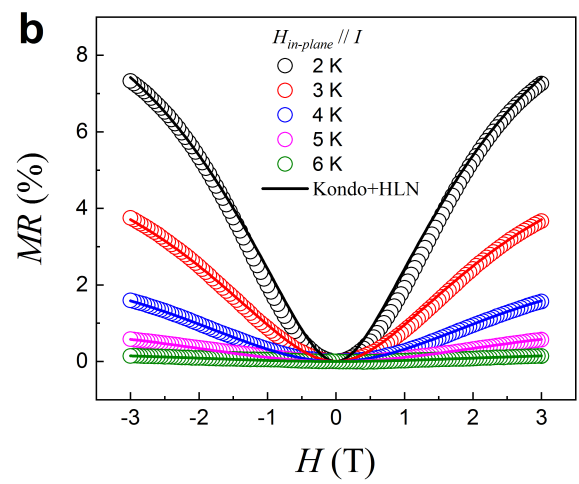

**Supplementary Fig. 14| Positive in-plane transverse MR$_{//}$ and longitudinal MR'$_{//}$ of the representative Nd$_{0.88}$Sr$_{0.12}$NiO$_2$ film, fitted by Kondo + HLN.** MR$_{//}$ ($H // $ ab, $H \perp I$) at temperatures of 2–6 K (**a**) and MR'$_{//}$ ($H //$ ab, $H // I$) at temperatures of 2–6 K (**b**). The solid curves are fitting to the Kondo + HLN.

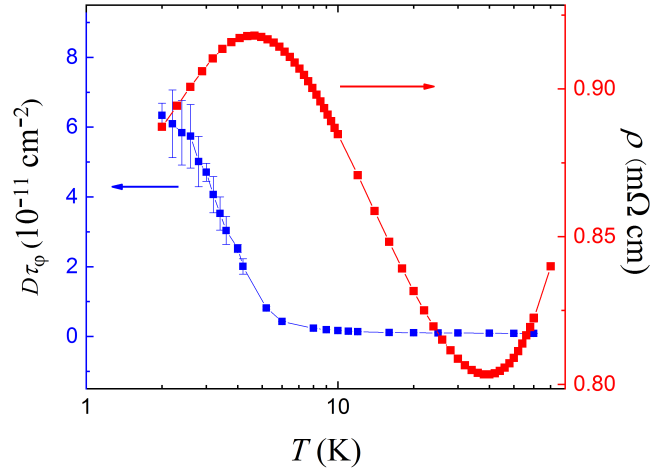

**Supplementary Fig. 15| Logarithmic temperature dependence of the phase coherence time $D \cdot \tau_\varphi$ (blue) and the resistivity (red) of the representative Nd$_{0.88}$Sr$_{0.12}$NiO$_2$ film at low temperatures.** It is interesting to plot the phase breaking length $l_\varphi^2 = D \cdot \tau_\varphi$ as a function of temperature on a logarithmic scale. It is also noteworthy that in this work the dephasing length l$_\varphi$ extracted from the fit is about 20–80 nm, much longer than the film thickness. Therefore, the phase coherent transport is two dimensional, justifying the application of the HLN equation for the data analysis. Taking *m*$^*$~ *20m$_e$* according to the measurements of nickelate bulks [54], the calculated $D$ is about ~10$^{-4}$ m$^2$/s, and thus estimated $\tau_\varphi$ is in the order of ~10$^{-11}$ s, which is consistent with that of dilute Kondo system [41]. The phase coherence time $\tau_\varphi$ shows a similar logarithmic temperature dependence as the resistivity, but the temperature range is slightly lower than that of the resistivity, the same behavior was also observed in the dilute Kondo system [55].

## Supplementary Note 1: *lnT*-type conductance correction by weak-localization/weak-antilocalization effect and electron-electron interaction

In two-dimensional system, the quantum interference correction on conductivity in a spin-orbital coupling system gives [34]:

$$\Delta\sigma_{WL/WAL} = \alpha p \frac{e^2}{\pi h} ln\frac{T}{T_0}. \tag{SN1}$$

Here, $T_0$ denotes the cut-off energy, $p$ is the dephasing exponent $\tau_\varphi^{-1} \sim T^p$, and $\alpha$ is an effective coefficient originating from the competition of weak-localization (WL) and weak-antilocalization (WAL) effect, with $\alpha > 0$ representing the WL correction and $\alpha < 0$ representing the WAL correction. Moreover, the combination effect of electron-electron interaction (EEI) and impurity scattering can further influence the density of states and induce additional conductivity correction [34]:

$$\Delta\sigma_{EEI} = \frac{e^2}{\pi h}\left(1 - \frac{3}{4}\tilde{F}\right) ln\frac{T}{T_1}. \tag{SN2}$$

Here, $T_1 = \frac{\hbar}{k_B \tau_0}$ denotes the thermal energy scale corresponding to impurity scattering with $\tau_0$ representing the mean free time, $\tilde{F}$ is commonly a positive constant determined by microscopic mechanisms. Thus, the total contribution from WL/WAL and EEI corrections reads:

$$\Delta\sigma_{EEI+WL/WAL} = \left(1 + \alpha p - \frac{3}{4}\tilde{F}\right)\frac{e^2}{\pi h} ln\frac{T}{T_0}. \tag{SN3}$$

As shown in Fig. 2c in the main text, $\alpha p \approx 0.213$, thus the total value $1 + \alpha p - \frac{3}{4}\tilde{F}$ is much smaller that the value $\beta = 6.96$. To summarize, in temperature range of 8-24 K with $lnT$ resistivity behavior, although the WL/WAL and EEI effect also contribute to the $lnT$ resistivity, the contribution from the Kondo scattering is dominant.

## Supplementary Note 2: Temperature dependence of resistivity

In the main text, we use the Hamann model to fit the measured ρ(T) curves for two reasons: 1) due to the onset of superconductivity, we cannot fit the saturation of resistivity below 7 K; and 2) the Hamann model can provide more useful information, such as magnetic impurity concentration. Three underdoped $Nd_{1-x}Sr_xNiO_2$ samples (x = 0, 0.05, 0.09). The main results are two-fold. Firstly, all three underdoped samples are well consistent with the Kondo scattering scenario down to 0.04 K. The Hamman fitting (Supplementary Fig. 5) provide quantitative justification for this point. Moreover, all the underdoped samples demonstrate clear feature of Kondo scattering in temperature regime above $T_K$. Secondly, the magnetic field truly influence the *R-T* curves of underdoped samples with pronounced negative magnetoresistance (Supplementary Fig. 5), reminiscent of the well-known negative magnetoresistance in Kondo system (La, Ce)Al$_2$ [31]. Under magnetic field, the magnetoresistance can be obtained by the modified Hamman model [31] by:

$$\rho_K(T/T_K) = \rho_0 \left\{1 - \frac{\ln(T/T_K)}{[\ln^2(T/T_K) + s(s+1)\pi^2]^{-1/2}}\right\} \cdot \left[1 - F^2\left(\frac{g\mu_B H}{k_B(T+T_K)}\right)\right]. \tag{SN4}$$

Here, F(x) is the Brillouin function for spin *s* and *g* denotes the Landé *g*-factor. Theoretical fitting for the *R-T* curves under magnetic field (as shown in Supplementary Fig. 5) by the modified Hamman mode also provide quantitative verifications for the Kondo scenario. Moreover, all three underdoped samples demonstrate the Fermi liquid behavior at low temperature below $T_K$.

## Supplementary Note 3: Magnetoresistance fitted by Kondo model

In this work, we used the HLN model to fit the measured magnetoresistance (MR) curves, mainly because the HLN model can provide more useful information, such as the dephasing field $H_\varphi$ and the fitting parameter α. To compare the MR with the theory, we can also use the zero temperature Kondo MR expression of Lee et. al. [56]. At zero temperature, a Kondo impurity magnetization is given as [56-58]:

$$M(H/H_1) = \begin{cases} \frac{1}{\sqrt{2\pi}} \sum_{k=0}^{\infty} \left(-\frac{1}{2}\right)^k (k!)^{-1} \left(k+\frac{1}{2}\right)^{\left(k-\frac{1}{2}\right)} e^{-\left(k+\frac{1}{2}\right)} \left(\frac{H}{H_1}\right)^{2k+1}, & H \leq \sqrt{2}H_1 \\ 1 - \pi^{-\frac{3}{2}} \int \frac{dt}{t} sin(\pi t) e^{-t \cdot ln\left(\frac{t}{2e}\right)} \left(\frac{H}{H_1}\right) \Gamma\left(t+\frac{1}{2}\right), & \sqrt{2}H_1 \leq H \end{cases} \quad (SN5)$$

where $H_1$ is a magnetic field scale, related to both Kondo temperature and the g-factor of impurity spin. We chose $H_1$ =11.0 T, according to the isomagnetics ρ(T) observation (see Fig. 1c). After calculating the magnetization of impurity, the zero temperature Kondo-magnetoresistance is given by:

$$R_K\left(\frac{H}{H_1}\right) = R_K(H=0) cos^2\left(\frac{\pi}{2} M\left(\frac{H}{H_1}\right)\right). \quad (SN6)$$

And the MR is given by

$$R(H) = R_0 + R_K\left(\frac{H}{H_1}\right). \quad (SN7)$$

In Supplementary Fig. 13, the solid lines are the fits with the experimental negative MR data in a temperature range of 10–20 K by Eqs. SN5, SN6 and SN7. The excellent fit between experimental data and theory points towards Kondo type scattering in this temperature range.

At T< 7 K, the positive MR appears to be the contribution of the weak anti-localization (WAL) and/or the onset of superconductivity. Thus, we add the WAL term by using HLN equation to the Kondo interaction term in MR equation and it follows as:

$$R(H) = R_0 + R_K\left(\frac{H}{H_1}\right) - \frac{\alpha e^2}{\pi h} R_0^2 \left[\Psi\left(\frac{1}{2}+\frac{H_\varphi}{H}\right) - ln\frac{H_\varphi}{H}\right]. \quad (SN8)$$

In Supplementary Fig. 14, the solid lines are the fitting curves considering both Kondo scattering and HLN formula with the experimental positive MR data in a temperature range of 2–6K by Eq. SN8. The deviation at low temperatures and higher magnetic fields is possibly due to the onset of superconductivity. The quality of the fit strongly suggests that the WAL effect or superconductivity overrides the Kondo scattering at T < 7 K.